\documentclass{article}[letterpaper,10pt]
\twocolumn
\usepackage[english]{babel}
\usepackage[noblocks]{authblk}
\usepackage{comment}
\usepackage{biblatex} 
\usepackage{float}
\usepackage[letterpaper,top=2cm,bottom=2cm,left=2cm,right=2cm,marginparwidth=1cm]{geometry}

\usepackage{amsmath}
\usepackage{graphicx}

\usepackage[colorlinks=true, allcolors=blue]{hyperref}

\title{High Fidelity Artificial Quantum Thermal State Generation using Encoded Coherent States}
\author[1,2,*]{Haley Weinstein}
\author[2,3]{Bruno Avritzer}
\author[2,3]{Todd A. Brun}
\author[1,2]{Jonathan L. Habif}

\affil[1]{Information Sciences Institute, University of Southern California, 4676 Admiralty Way, Marina Del Rey, CA 90292, USA}
\affil[2]{Ming Hsieh Department of Electrical and Computer Engineering, University of Southern California, 3740 McClintock Ave Suite 100, Los Angeles, CA 90089, USA}
\affil[3]{Department of Physics and Astronomy, University of Southern California,  825 Bloom Walk, Los Angeles, CA 90089, USA}

\affil[*]{haweinst@usc.edu}

\begin{document}
\maketitle

\begin{abstract}
Quantum steganography is a powerful method for information security where communications between a sender and receiver are disguised as naturally occurring noise in a channel. We encoded the phase and amplitude of weak coherent laser states such that a third party monitoring the communications channel, measuring the flow of optical states through the channel, would see an amalgamation of states indistinguishable from thermal noise light. Using quantum state tomography, we experimentally reconstructed the density matrices for artificially engineered thermal states and spontaneous emission from an optical amplifier and verified a state fidelity $F>0.98$ when compared with theoretical thermal states.
\end{abstract}

\section{Introduction}
\begin{figure*}
\centering\includegraphics[width=\textwidth]{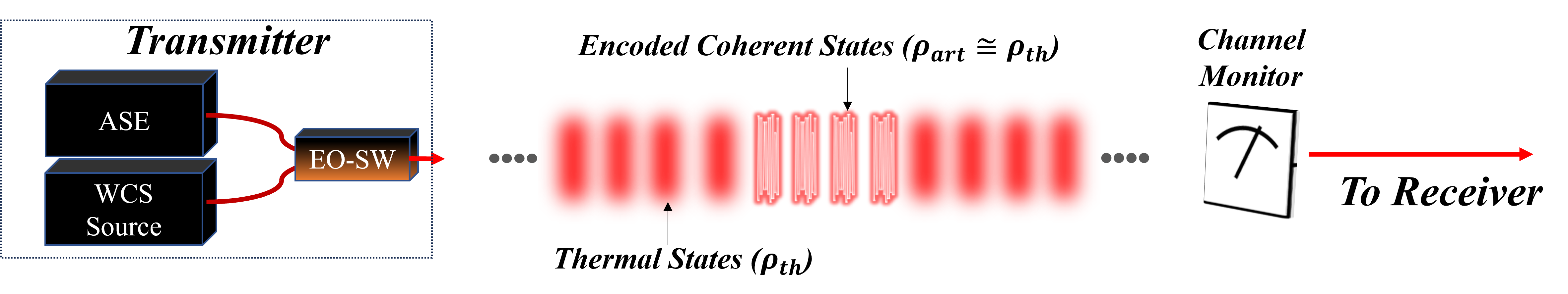}
\caption{A channel with amplified spontaneous emission (ASE) contains sideband, thermal state noise represented by a thermal quantum density matrix $\rho_{th}$. A phase and intensity modulated weak coherent source (WCS) generates the quantum density matrix $\rho_{art}$ which mimics a quantum thermal state. Either $\rho_{th}$  or $\rho_{art}$ is then selected by an electro-optic switch (EO-SW). A channel monitor would be expecting to measure quantum thermal states due to the amplified spontaneous emission. Since $\rho_{art} \sim \rho_{th}$ the channel monitor would not be able to detect communications. However, since the desired receiver shares information with the transmitter, they could measure data from the coherent encoded parts of the signal.}
\label{fig1diagram}
\end{figure*}
High fidelity quantum states are essential resources for quantum computing, communications, and sensing, as the performance of any quantum protocol or algorithm relies heavily on the capability to prepare the quantum state required for the protocol with high precision.  An exciting topic in the field of optical communications has emerged examining the fundamental quantum limits to the covertness of a communications or sensing channel when a third-party eavesdropper possesses the capabilities of quantum measurement and uses them as resources to discover whether or not communications or sensing is occurring in the channel\cite{squarerootlaw,bash2013limits}.  Covert communications, also referred to as Low Probability of Detect (LPD) communications, considers a transmitter and receiver communicating over a channel monitored by a third party tasked with identifying the mere presence of communications.  The objective of the transmitter and receiver is to mask the physical encoding of the communications signal such that it appears innocuous on the channel.  A technique for masking the communications signal is designing it to mimic, as closely as possible, the original channel statistics presented to the passive channel monitor by the quiescent noise in the channel.  

Key results from quantum-based analyses of LPD communications have shown two distinct operating cases. When the sender and receiver are communicating through a channel with naturally occurring noise, equally random to all three parties, $N$ uses of a channel with additive white Gaussian noise (AWGN) results in a number of bits proportional to \begin{math} \sqrt{N} \end{math} that can be communicated covertly.  This result was proven theoretically in \cite{squarerootlaw} and validated experimentally in \cite{bash2015quantum}.  In this case a finite number of bits can be successfully transmitted with a bounded covertness, while the asymptotic communication rate is 0.  Alternatively, when the sender and receiver are communicating through a channel with noise that can either be modulated by the sender, or the statistics of which are well known to the sender and receiver, then provably covert positive-rate communications can be achieved\cite{soltani2018covert}.  

As steganography is a technique to achieve covert communications by hiding messages within seemingly benign larger messages \cite{singh2000code}, quantum steganography was originally proposed for disguising communication symbols as incoherent errors on qubits on a quantum channel \cite{shaw2011quantum}.  Congruent with the analyses in \cite{squarerootlaw} and \cite{soltani2018covert} analyses for quantum steganography were conducted in a channel with and without environmental noise \cite{sutherland2020quantum,sutherland2019quantum}. Recently, these results were adapted to implementations using quantum states in an optical channel \cite{avritzer2024quantum}.

Lasers and optical amplifiers are essential ingredients in optical communications transmitters but bring the burden of broadband, amplified spontaneous emission (ASE) noise to the channel.  A quantum steganography protocol for covert communications can leverage this ever-present noise by hiding a communication transmission in the noise so that, to a third-party monitoring the channel, the ASE and the communications signal are indistinguishable. For example, in \cite{Wu:13} authors demonstrate an optical steganography experiment where ASE noise is itself modulated and used as a carrier for covert information.  

To carefully mimic the quantum state of a single spatio-temporal mode of ASE our approach is to filter a narrow optical bandwidth of ASE from the channel and replace it with coherent state symbols encoded according to the prescription described in \cite{avritzer2024quantum}; this process is shown in Fig. \ref{fig1diagram}. This approach prevents even an eavesdropper with quantum measurement capabilities from detecting communications, discriminating information carrying states from the quantum states generated by ASE. This band chosen to implement quantum steganographic communications can be selected out of band from strong active laser signals used to carry traditional non-covert optical traffic.  When communications commence this narrow band of ASE is filtered out of the channel and replaced with information-carrying coherent state light with coding in amplitude and phase such that the third party monitor is unaware to any disruption in the channel, but a receiver, aware of the coding approach and the arrival time of the message, can successfully decoded the information from the transmitter.  The third-party channel monitor can make quantum measurements on the states passing through the channel.  As these are low intensity states, however, quantum mechanics limits the amount of information that can be extracted with a single measurement or measurements over many copies of a quantum state, quantified by the Helstrom bound and the quantum Chernoff bound, respectively \cite{Habif:21,jagannathan2022demonstration}. 

In this paper we describe our approach to tailoring optical communications symbols in coherent state amplitude and phase, following the prescription in \cite{avritzer2024quantum}, such that a sequence of measurements on the states renders them indistinguishable from thermal states. Custom constellation shaping has found application in high spectral efficiency optical communications to achieve energy efficient constellation maps that are more robust against noise statistics in a channel \cite{buchali2015rate}.  In a similar strategy, our objective is to generate a constellation map for coherent state symbols such that the quantum state of the transmitted symbols (which is a mixed state) closely mimics the quantum state of the thermal noise channel.  

Using optical fiber-based components, we report a demonstration creating a statistical mixture of weak coherent laser states closely mimicking a quantum mechanical thermal state: a maximally mixed state with Bose-Einstein distributed photon number statistics and completely randomized phase.  These states were engineered to match quantum thermal states with a mean photon number $\bar{n} \sim1.5$ photon per temporal mode from a narrow-band selection of ASE from an erbium-doped fiber amplifier. To quantitatively measure the quantum state fidelity ($F$) between our \emph{artificial} thermal state ($\rho_{art}$) and the naturally generated thermal state from the EDFA ($\rho_{th}$) we implemented a quantum state tomography measurement system to experimentally reconstruct the density matrices ($\hat{\rho}_{art}$ and $\hat{\rho}_{th}$) from our quantum state generator.  We compared the reconstructed matrices with a theoretical thermal state and found a very high reconstruction fidelity $ F(\rho_{th}, \hat{\rho}_{th})>0.98, F(\rho_{th}, \hat{\rho}_{art})>0.98$, additionally we found a high fidelity when comparing the experimentally reconstructed density matrices ($F\left(\hat{\rho}_{art},\hat{\rho}_{th} \right)>0.97$). 
\section{Theory and Numerical Results}
Lasers emit quantum mechanical pure coherent states $|\alpha\rangle, \alpha=|\alpha|e^{i\theta}$, that can be encoded with a prescribed distribution of amplitudes $|\alpha|$ and phases $\theta$. As shown in \cite{quantum_optics}, the thermal state can be described in this phase space as in the Glauber $P$ Representation. Additionally, the coherent state basis is over-complete, so the existence of a coherent-state representation is guaranteed \cite{quantum_optics}. This is shown by the following representation of the thermal state:
\begin{equation}
    \rho_{th} = \sum_{n=0}^\infty \frac{\bar{n}^{n}}{\left(\bar{n}+1\right)^{n+1}} |n\rangle \langle n| = \frac{1}{\pi\bar{n}}\int d^{2}\alpha e^{-c|\alpha|^{2}} | \alpha\rangle \langle \alpha |.
    \label{eq:thermalstate}
\end{equation}

In eq. \ref{eq:thermalstate} $|n\rangle$ is the photon Fock basis, $\bar{n}$ is the mean photon number in a single, spatio-temporal mode of the field and $c = 1/\bar{n}$. This representation shows the thermal state can be created as a statistical mixture of coherent states with uniformly distributed $\theta$, and field amplitudes $|\alpha|$ following a Rayleigh distribution:
\begin{equation}
Rayleigh (|\alpha|, \sqrt{\frac{\bar{n}}{2}})=\frac{2}{\bar{n}}|\alpha|e^{-\frac{|\alpha|^2}{\bar{n}}}.
\label{rayl}
\end{equation} 
In practice, we construct the artificial thermal state $\rho_{art}$ as a discrete summation over a set of coherent states using $M=L\cdot Q$ randomly chosen combinations of amplitudes and phases where the amplitudes and phases are selected from a set:$\{|\alpha|_l\}$ with $L$ elements and $\{\theta_q\}$ with $Q$ elements, respectively. In this form we write,
\begin{equation}
    \rho_{art}=\sum_{l=1}^L\sum_{q=1}^Qp_{q, l} |\alpha_{l, q}\rangle\langle\alpha_{l, q}|
\end{equation}
 where $p_{q, l}$ is the probability of each coherent state  \begin{equation}
    |\alpha_{l, q}\rangle=\sum_n|\alpha|_l^ne^{i\theta_q n}\frac{e^{-|\alpha|_l^2/2}}{\sqrt{n!}}|n\rangle.
\end{equation}
 participating in the mixture being generated.  These coefficients $p_{q, l}$ are carefully computed such that $\rho_{art}$ reproduces $\rho_{th}$ as faithfully as possible. 

To determine the reliability with which our engineered $\rho_{art}$ recreates $\rho_{th}$ we calculate three quantum information theoretic quantities: the quantum state fidelity ($F$), the Helstrom bound for quantum state discrimination ($ P_{\varepsilon}$), and the Von Neumann Entropy ($S(\rho)$).   The quantum state fidelity gives a measure of "closeness" between two quantum states and is calculated as: 
\begin{equation}
F(\rho_{th}, \rho_{art}) = Tr \left(\sqrt{\sqrt{\rho_{art}}\rho_{th}\sqrt{\rho_{art}}}\right)^2
 \label{fidelity}
\end{equation}
where $Tr(\cdot )$ denotes the trace operator.  This number quantifies the closeness of the two matrices, with $F = 1$ indicating the matrices are identical.
The Helstrom Bound defines a lower bound on the minimum probability of error for discrimination between quantum states achievable with \textbf{any} measurement strategy:
\begin{equation}
    P_{\varepsilon} \geq \frac{1}{2}-\frac{1}{4}||\rho_{th}-\rho_{art}||_1.
    \label{helstrom}
\end{equation}
The von Neumann entropy is the quantum extension of Gibbs entropy \cite{entropy}. For a given quantum system the von Neumann entropy is defined as
\begin{equation}
    S(\rho) = -Tr(\rho log_2(\rho))
    \label{ent}
\end{equation}
and quantifies the mixedness of a state. The von Neumann entropy of a thermal state is maximal whereas the von Neumann entropy of a pure, coherent state is 0. 

 In order to decide how many samples, $M$, is necessary for a high fidelity reconstruction we numerically calculated Eq. \ref{fidelity} for different values of $M$. In these calculations we assumed the number of discrete intensity samples, $L$, and phase samples, $Q$, were equal.
\begin{figure}[H]
\centering\includegraphics[width=10cm]{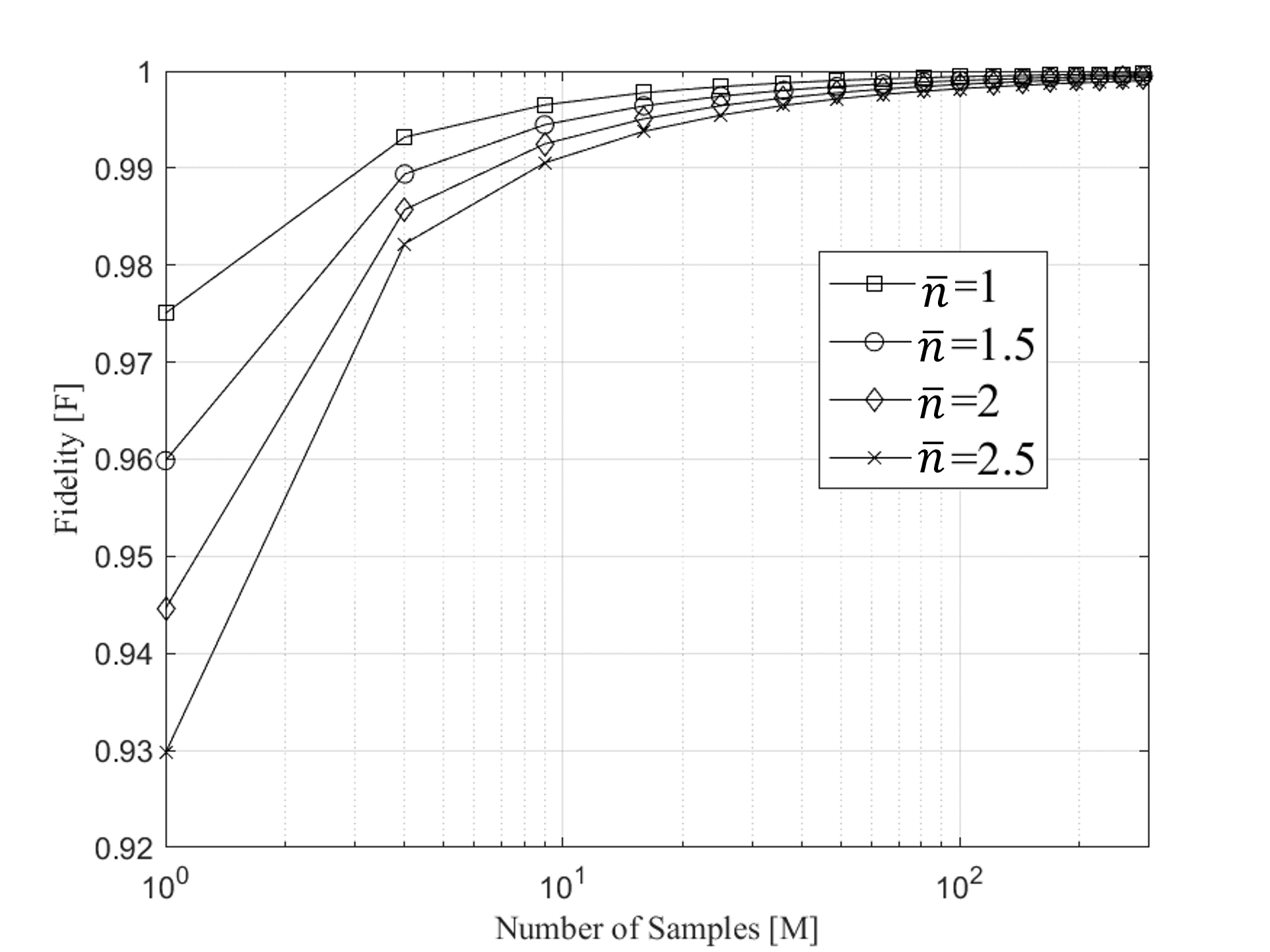}
\caption{Calculated fidelity ($F$) from Eq. \ref{fidelity} plotted for different number of samples ($M$) and mean photon numbers ($\bar{n}$)}
\label{simresults}
\end{figure}
As shown in Fig. \ref{simresults}, a number of samples greater than 50 will yield a $\rho_{art}$ which has a  $F\geq0.99$ for a range of mean photon numbers. Thermal states with higher mean photon numbers require a higher number of samples to make a high-fidelity reconstruction. For this experiment we selected a number far above this. Different levels of covertness, and different communication scenarios could call for a different number of samples. 

\section{Experimental Design}
\begin{figure*}
\centering\includegraphics[width=\textwidth]{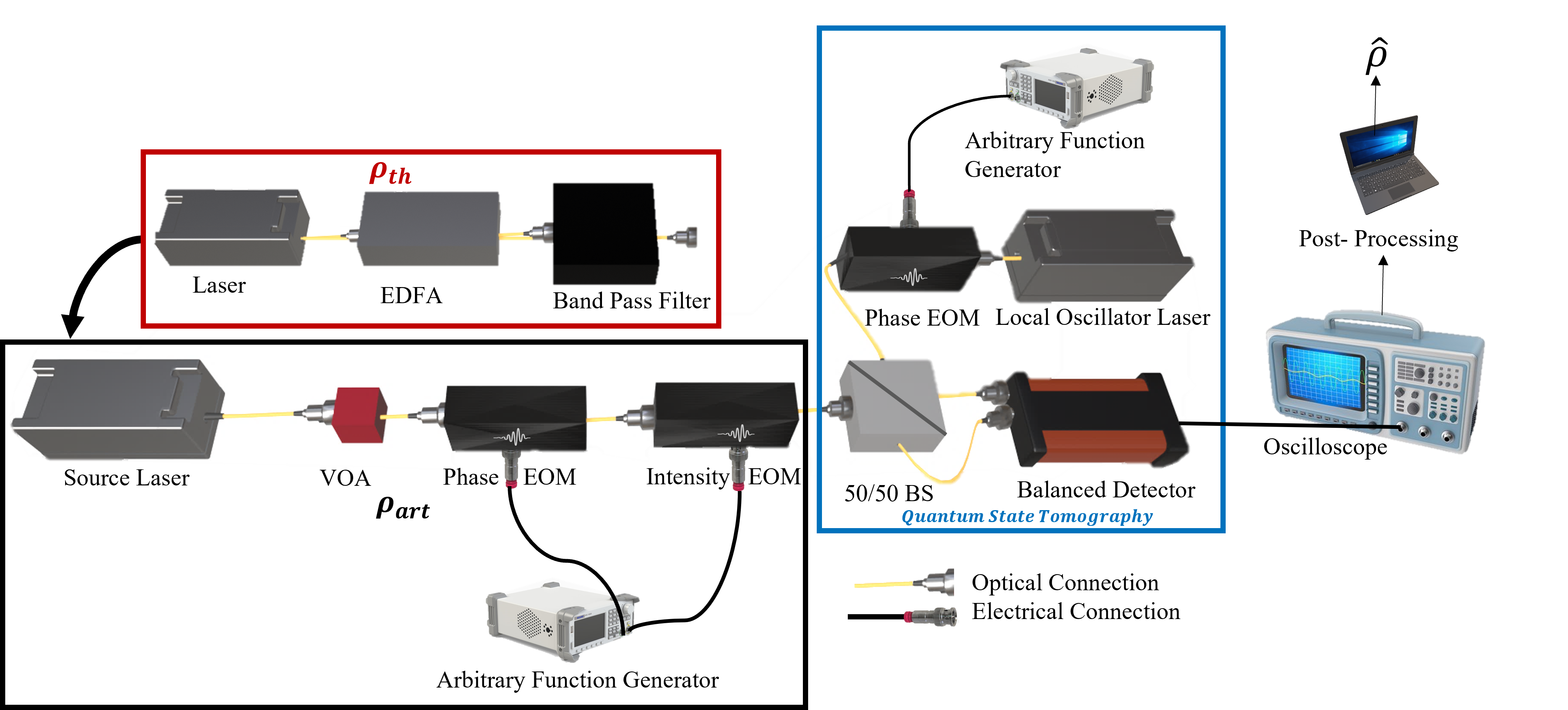}
\caption{Experimental setup. Diagram for generating real thermal states is highlighted in a red box ($\rho_{th}$). Diagram for creating artificial thermal states, generated by coherent state encoding, is highlighted in a back box($\rho_{art}$). Quantum State Tomography optics and electronics are highlighted in a blue box.}
\label{qst}
\end{figure*}
To generate the artificial thermal states, we must generate then engineer the photostatistics of coherent states. We go beyond previous thermal state generation techniques such as \cite{Straka:18} and account for coherent detection techniques. To generate the coherent states, we use a shot noise limited, low linewidth laser, the RIO Orion module at 1560.625 nm. We attenuate the laser using a manual variable optical attenuator (VOA) from Agiltron so that the mean photon number mimics the mean photon number of the desired thermal state. To engineer the photostatistics of the coherent states, electro-optic modulators (EOM) are used to impose an intensity and phase. Both EOMs are purchased from Thorlabs. The intensity EOM, part number LN81S-FC has a bandwidth of 10 GHz. This intensity EOM has a measured optical extinction ratio of $\sim$ 25 dB. To ensure this was enough dynamic range, we converted the discretized photon number values($\bar{n}_m$) shown in Fig. \ref{simresults} to desired output power, $P_m$, values within a 13 ns temporal mode ($\tau$) using Planck's constant ($h$) and the optical frequency ($f$):
\begin{equation} P_m = \frac{\bar{n}_{m}hf}{\tau}.\end{equation}

We found the range of powers was well within the 25 dB dynamic range of the intensity EOM.  The phase EOM, part number, LNP4216 has an operational bandwidth up to 40 GHz. This phase EOM has a low $V_\pi$ value so that within the specifications of the EOM we can take advantage of the entire 0 to 2$\pi$ dynamic range. Both modulators are driven by an Arbitrary Function Generator, the Siglent SDG6052X, with a maximum operable bandwidth of 500 MHz. The phase and intensity are modulated well within these specifications. In practice, we modulate the phase and intensity at the same bandwidth of our detector, 75 MHz, in order to remain in one temporal mode. This setup is highlighted in a black box and depicted in Fig. \ref{qst}. Ultimately, these intensity and phase values will be used to create a communications codebook. 

To generate real thermal states, another laser with a slightly different center wavelength, also in the C-Band (1530-1565nm), is amplified by an EDFA. The laser is a diode laser from Thorlabs, part number LPS-1550-FC, and the EDFA is single mode booster from Agiltron, part number EDFA-1C2111333. A tunable bandpass filter from Newport, part number TBF-1550-1.0-FCAPC,  carves out 100 GHz of thermal noise caused by ASE around the center frequency of the local oscillator as shown in Fig.\ref{qst}.

To perform quantum state tomography on either the real or artificial thermal states, the phase of a strong local oscillator is swept from 0 to 2$\pi$ and is used to perform balanced homodyne detection. The local oscillator is generated by the same laser as is the artificial thermal state. The power of the local oscillator was 100$\mu$W. We performed vacuum state balanced homodyne measurements and calculated the scaled variance of the result to ensure that this power was within the shot noise limit. When measuring the artificial thermal state, a small portion of this laser is split off via a beamsplitter to use as the signal source. The signal source is approximately 8 orders of magnitude smaller than the local oscillator at the detector. The phase of the local oscillator is then modulated via another LNP4216, low $V_{\pi}$, phase EOM. A Tektronix AFG1062 drives the phase EOM to 50 equally spaced phases between 0 and 2$\pi$ for 1 ms at each phase. Quantum state tomography is implemented experimentally when the signal (real or artificial) and phase modulated local oscillator are mixed through a 50/50 beamsplitter, and the resulting two modes are directed to a Thorlabs PDB425C balanced detector where the light is detected by two photodiodes and the resultant difference current from these photodiodes is amplified. The PDB425C balanced detector has a high, >35 dB common mode rejection ratio (CMRR). This leads to low electronic noise, which is critical in maintaining a shot noise limited measurement.

The balanced detectors radio frequency output is measured at a Tektronix MDO34 oscilloscope, leading to a dataset of voltages and the corresponding phase of the local oscillator when that voltage was measured: ($V_k$, $\theta_k$). These measurements are transformed into the measured state's quadrature probability function by normalizing the detected variance and amplitude to a measured vacuum state, as introduced in \cite{ORNL}: 
\begin{equation}
    x_{\theta,k} =(V_k-V_{Vac})\sqrt{1/(4\cdot \sigma_{Vac}^2)}
    \label{quadrature}
\end{equation}
where $V_{Vac}$ and $\sigma_{Vac}$ are the measured voltage and standard deviation when the signal input is a vacuum state. The variance of a vacuum state in balanced homodyne detection operating at the shot noise limit is $1/4$ \cite{Shapiro2009TheQT}, leading to the scaling factor in Eq. \ref{quadrature}. Using the Maximum Likelihood Estimator shown in \cite{mle}, we recreate the Fock basis density matrices from the measured quadrature probability functions. The likelihood function to maximize is the product of the probability density functions for each quadrature measurement given the input local oscillator phase: 
\begin{equation}
    L(\rho) = \prod_{k=1}^K\sum_{m=0}^{n_c}\sum_{n=0}^{n_c}\rho_{mn}\frac{e^{i(n-m)\theta_k}}{\sqrt{\pi m!n!2^{m+n}}}e^{-x_k^2}H_m(x_k)H_n(x_k).
\end{equation}
where $H_n(x)$ is the Hermite polynomial of variable $x$ of the $n^{th}$ order and $n_{c}$ is the cutoff photon number, chosen for numerical conversion. 
\section{Results and Discussion}
\begin{figure*}
\centering
\includegraphics[width=\textwidth]{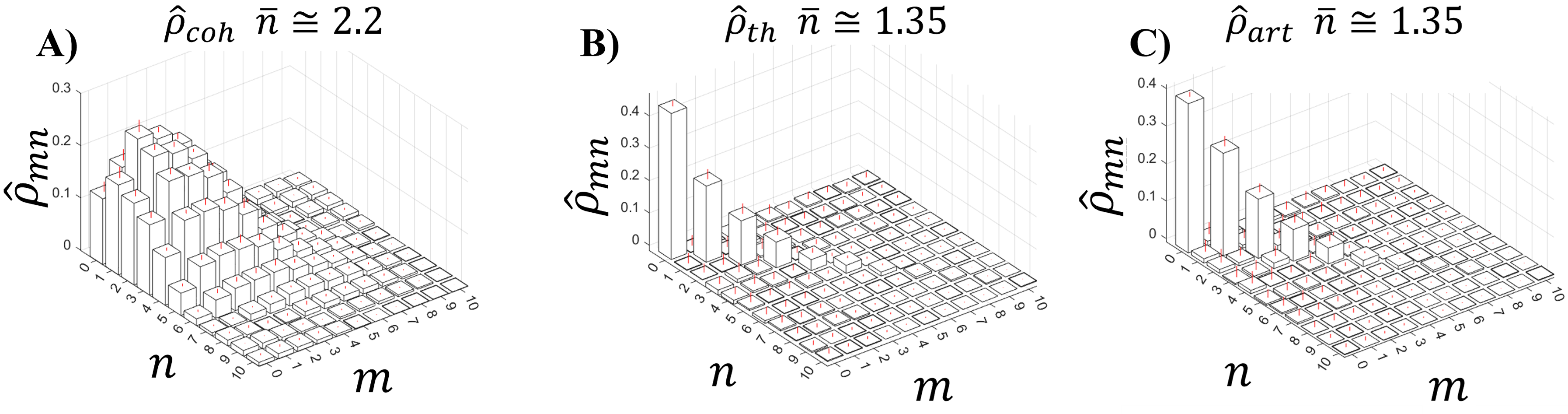}
\caption{Graphical representations of density matrices reconstructed from measurements by Quantum State Tomography. Graphs presented for the coherent states (A) used to mimic the naturally occurring thermal states (B) to create artificial thermal states (C). The x axis and y axis represent the row and column respectively of the reconstructed density matrix. The losses induced by the modulators for coherent encoding causes a reduction in the mean photon number from (A) to (C). Each reported mean photon number is calculated from the averaged density matrix reconstruction.}
\label{data}
\end{figure*}

We experimentally reconstructed the density matrices for the coherent states (Fig. \ref{data}A), the artificially generated thermal states (Fig. \ref{data}C ), and the  states from the EDFA (Fig. \ref{data}B). Using Eq. \ref{quadrature} we calculated 2000 quadrature elements to form a reconstructed density matrix. Then, we averaged over 10 density matrices and plotted the real part of the resultant estimated density matrix with error bars on matrix elements equal to $\pm$1 standard deviation.  

The fidelity of each of the experimentally reconstructed density matrices was calculated with respect to the theoretical thermal state density matrix, yielding 
\begin{equation} F(\rho_{th}, \hat{\rho}_{th})>0.98, F(\rho_{th}, \hat{\rho}_{art})>0.98. \label{fi1} \end{equation}
Direct comparison between the two experimentally reconstructed density matrices also yielded high quantum state fidelity
\begin{equation}  F(\hat{\rho}_{th}, \hat{\rho}_{art})>0.97. \label{fi2} \end{equation} We found similarly high-fidelity results for mean photon numbers in the range $1<\bar{n}<2$. The fidelity results in Eq.\ref{fi1} and \ref{fi2} verify our coherent encoded state is nearly indistinguishable from both a theoretical thermal state and measured thermal state. This measurement goes beyond classical or semi-classical measurements, such as \cite{Straka:18}, as it considers the entirety of the quantum density matrix, not just diagonal elements. Such an analysis is necessary to understand the vulnerability of a communications protocol to an eavesdropper with arbitrary quantum measurement capability. Additionally, we used Eq. \ref{ent} to calculate the von Neumann entropy of the reconstructed and theoretical states. We found $S(\hat{\rho}_{art})=2.12$ which is near the theoretical maximum for the given mean photon number, $S(\rho_{th}) = 2.31$. In a future work, we will engineer the photo-statistics of the coherent states with a set communication codebook such that the time correlation of the engineered states is as close as possible to the real thermal states. If we consider a Lindblad description of the EDFA with a number of emitters coupled to an output waveguide, it is possible to generate a time-dependent solution for the density matrix of the emitted light.

Using these reconstructed density matrices we can also compute the eavesdropper's quantum limit to discrimination error probability between the EDFA states and the artificially generated thermal states with the Helstrom bound, $P_{\varepsilon}(\hat{\rho}_{th}, \hat{\rho}_{art})\geq 0.45$.  This discrimination error probability well exceeds the Helstrom bound reported $P_{\varepsilon} \sim 0.14$ in \cite{Habif:21} for discriminating between thermal and laser states, with $\bar{n}\sim 1$.  We believe that experimental non-idealities, such as fluctuating source intensity, are responsible for $P_{\varepsilon}(\hat{\rho}_{th}, \hat{\rho}_{art})<0.5$ which can be addressed with further precision in engineering of the source generating $\rho_{art}$.

The results of this experimental work demonstrate a proof of concept validating the covert communications approach described in \cite{avritzer2024quantum} paving the way towards a full implementation of the protocol.  The full density matrix reconstruction is essential to determine the capability of a quantum-capable eavesdropper for detecting the presence of communications in a noisy channel.   Importantly, experimental quantification of quantum information theoretic quantities, such as the Helstrom bound, provide insight into the limitations of direct encoding of single mode coherent states to achieve covertness for LPD communications. This motivates investigation into enhanced techniques to even more accurately mimic the quantum state of light generated by amplified spontaneous emission, or other thermal sources.  Finally, as with any information security research, it will be interesting and important to investigate the capabilities of an eavesdropper, devising a strategy for realizing a structured receiver architecture that can approach or achieve the Helstrom bound for discrimination reported herein.
\section{Funding} This material is based upon work supported by the Defense Advanced Research Projects Agency (DARPA) under Contract No. HR001124C0403 and NSF Grants 1719778, 1911089 and 2316713.

\section{Acknowledgments} The authors are grateful to Dr. Joseph Chapman and team for their willingness to share their insights and expertise. 

\smallskip

\section{Disclosures} The authors declare no conflicts of interest. Any opinions, findings and conclusions or recommendations expressed in this material are those of the author(s) and do not necessarily reflect the views of the Defense Advanced Research Projects Agency (DARPA).

\section{Data availability} Data underlying the results presented in this paper are not publicly available at this time but may be obtained from the authors upon reasonable request.



\printbibliography

\end{document}